\documentclass[useAMS]{mn2e}
\usepackage{lscape}
\usepackage{rotating}
\usepackage{amssymb}
\usepackage{float}
\usepackage{hyperref}
\usepackage{newtxtext,newtxmath}

\def\msun{\hbox{M$_\odot$}}

\def\vrot{\hbox{V$_{\rm surf}$}}

\def\t4{\hbox{t$_{\rm 4}$}}

\def\cm3{\hbox{cm$^{-3}$}}
\input psfig.sty
\input epsf.sty
\usepackage{graphicx}
\usepackage{epsfig}
\title[Rotation in PMS stars]
{On the Origin of the Bimodal Rotational Velocity Distribution in Stellar Clusters:  Rotation on the Pre-Main Sequence}
\author[Bastian et al.] {Nate Bastian,$^1$ Sebastian Kamann,$^{1}$ Louis Amard,$^{2}$ Corinne Charbonnel,$^{3,4}$ \newauthor  Lionel Haemmerl\'e,$^{3}$ and Sean P. Matt$^{2}$\\
$^{1}$Astrophysics Research Institute, Liverpool John Moores University, 146 Brownlow Hill, Liverpool L3 5RF, UK\\
$^{2}$University of Exeter, Department of Physics \& Astronomy, Stoker Road, Devon, Exeter EX4 4QL, UK\\
$^{3}$ Department of Astronomy, University of Geneva, Chemin de P\'egase 51, 1290 Versoix, Switzerland\\
$^{4}$ IRAP, UMR 5277 CNRS and Universit\'e de Toulouse, 14, Av. E.Belin, 31400 Toulouse, France \\
}
\date{Accepted. Received; in original form}
\pagerange{\pageref{firstpage}--\pageref{lastpage}}
\pubyear{2020}
\begin{document}
\maketitle
\label{firstpage}
\begin{abstract}
We address the origin of the observed bimodal rotational distribution of stars in massive young and intermediate age stellar clusters.  This bimodality is seen as split main sequences at young ages and also has been recently directly observed in the $Vsini$ distribution of stars within massive young and intermediate age clusters.  Previous models have invoked binary interactions as the origin of this bimodality, although these models are unable to reproduce all of the observational constraints on the problem.  Here we suggest that such a bimodal rotational distribution is set up early within a cluster's life, i.e., within the first few Myr.  Observations show that the period distribution of low-mass ($\la 2 M_\odot$) pre-main sequence (PMS) stars is bimodal in many young open clusters and we present a series of models to show that if such a bimodality exists for stars on the PMS that it is expected to manifest as a bimodal rotational velocity (at fixed mass/luminosity) on the main sequence for stars with masses in excess of $\sim1.5$~\msun.  Such a bimodal period distribution of PMS stars may be caused by whether stars have lost (rapid rotators) or been able to retain (slow rotators) their circumstellar discs throughout their PMS lifetimes.  We conclude with a series of predictions for observables based on our model.
\end{abstract}

\section{Introduction}
\label{sec:intro}

Resolved young and intermediate age massive clusters in the Magellanic Clouds display a number of unexpected features in their colour magnitude diagrams (CMDs).  These include extended main sequence turnoffs (e.g., Mackey \& Broby Nielson~2007) and split, or dual, main sequences (e.g., Milone et al.~2016).  Both phenomena appear to be largely driven by the underlying distribution of stellar rotational velocities (e.g., Bastian \& de Mink~2009; D'Antona et al.~2015; Dupree et al.~2017; Kamann et al. 2018; 2020; Bastian et al. 2018; Marino et al. 2018; Georgy et al.~2019).  In particular, the split-MS is thought to be due to a bimodal rotational distribution with one peak being made up of slowly rotating stars and the other made up of rapid rotators (D'Antona et al. 2015), potentially near critical rotation (Bastian et al. 2017; Milone et al. 2018).  Such a bimodal rotational distribution has recently been observed in a massive, $\sim1.5$~Gyr cluster in the LMC (NGC~1846), with (Vsini) peaks at 60 and 150~km/s (Kamann et al. 2020).  Milone et al.~(2018) have found that the split MS occurs for stars with masses as low as $1.6$~\msun\  and as high as $\sim5$~\msun, already at an age of $40$~Myr, suggesting that it occurs from an early age.

This naturally raises the question of where such a bimodal rotational distribution could come from.  D'Antona et al.~(2015) have suggested that interacting binaries may play an important role, namely that if all/most stars are born as rapid rotators then interacting binaries could brake them, resulting in a population of slowly rotating stars.  This model is able to produce an extended main sequence turn-off (eMSTO) as well as a dual main sequence. % \sout{This model} 
It predicts a higher binary fraction amongst slow rotators than fast rotators, which is at odds with observations of the only cluster studied in this way to date, namely that the binary fraction is similar between the fast and slow rotators in NGC~1846 (Kamann et al.~2020).  Clearly, further studies are needed to confirm or refute this general behaviour.  Additionally, this model predicts that, because they should be predominantly made up of binary systems, the slowly rotating stars should be more centrally concentrated within the cluster.  This is due to the fact that, as binaries, they are on average higher mass than the rapidly rotating single stars, meaning that mass segregation will act on them (e.g., Hut et al.~1992).  Also, since the binary fraction of stars increases towards the cluster centre (e.g., Hurley et al.~2007; Milone et al.~2012; Giesers et al.~2019) we would expect more slowly rotating stars (in binaries) towards the cluster centres.  This is the opposite as seen in young massive clusters where the rapid rotators tend to be  more centrally concentrated (e.g., Milone et al.~2018) or the two populations have similar profiles (Li et al.~2017).

Similarly, a number of studies have explored the role of interacting binaries in causing eMSTOs,  particularly at younger ages ($<100$~Myr; e.g., D'Antona et al.~2017; Beasor et al.~2019; Wang et al.~2020).  These models make clear predictions as to the rate of binarity in different parts of the CMD (i.e. along the blue upper main sequence turnoff) that can be directly tested with observations.  However, as the interacting binary fraction increases with increasing stellar mass it is unlikely to play a major role in the observed eMSTOs and dual main sequences observed in clusters at older ages (i.e., for stars with masses below $\sim5$~\msun).  Finally, Ram{\'\i}rez-Agudelo et al. (2015), using data from the VLT FLAMES Tarantula Survey, found that in 30~Doradus, high mass single O-stars have a similar rotational distribution than the O-stars in binaries, suggesting that binarity does not govern the rotational distribution of (at least of high-mass) stars.

In the current work we present an alternative model for the origin of the bimodal rotational distribution in massive clusters where the distribution is set during the formation and early evolution of the cluster.  In \S~\ref{sec:model} we present the model and its implications and summarise our results and predictions in \S~\ref{sec:discussion}.

\section{Model}
\label{sec:model}

\subsection{The Early Period Distribution of Stars in Clusters}

%{\bf shift this section around - first just bimodal period distributions and then the possible causes} 
As an alternative to the binary based model of D'Antona et al.~(2015), the rotational velocity distribution could be set during the very early stages of cluster formation and stellar evolution.  
In particular, bimodal rotation period distributions are sometimes (although not always) seen for the low-mass stars in clusters with ages of only a few Myr.
For example, stars with masses greater than 0.4~$M_\odot$\footnote{The mass above which some cluster stars show a bimodal rotation distribution is sometimes given as 0.4, and sometimes 0.25~$M_\odot$, depending on the models used by the authors.} were shown to have bimodal rotation period distributions in the Orion Nebular Cluster (Herbst et al. 2001; 2002), NGC 2264 (Lamm et al. 2005), and IC 348 (Cieza \& Baliber 2006).  
See also compilations in Irwin \& Bouvier (2009) and Bouvier et al.\ (2014).
These observed distributions typically have small numbers of stars with greater than a solar mass, but the slightly older open cluster, hPer ($\sim13$~Myr), displays a clear bimodal period distribution, with peaks at $0.9$ and $7$~days in the full mass range of 0.3--1.4 $M_
\odot$ (Moraux et al.~2013).  The rotation of intermediate mass PMS stars, and Herbig Ae-Be rotation in particular, has been studied by Alecian et al.~(2013).  These authors show that magnetic Herbig stars are much slower rotators than their non-magnetic counterparts, from very early times up to a few million years. Although the fraction of magnetic stars is low in this mass range ($5-10$\%), and the mechanism responsible for their slow rotation rate is still quite uncertain, the rotation period bimodality is still present in this case.
%{\bf There is tentative evidence that this behaviour extends to higher stellar masses (up to at least $4$~\msun - Sabbi et al.~2020).  }
The origin of this bimodal period distribution is still not fully understood, but it is thought to be a manifestation of a star-disc interaction (SDI) during the PMS evolution (e.g., Bouvier et al.~1993).   We will discuss this in more detail in \S~\ref{sec:origin}.

%For example, the pre-main sequence stars in the $\sim2$~Myr-old Orion Nebular Cluster (ONC) show a clear bimodal rotational distribution with peaks at $\sim1.5$ and $\sim7.5$ days, at least for stars with masses in excess of $0.4$~\msun\ (Herbst et al. 2002). 
% These stars are still contracting towards the main-sequence, so unless a breaking mechanism is able to slow the star down it will rotate faster as it approaches the main sequence.  However, if a star has an accretion disc, 'disc-locking' may occur (magnetic coupling of the disc to the star) which acts as a drag force, slowing the star down (e.g., Bouvier et al.~1993; Rebull et al.~2004; Bouvier et al.~2014).  Indeed, there is evidence that within a given young cluster, stars without discs rotate more rapidly than stars with discs (e.g., Cieza \& Baliber~2007). 

%Bimodal period distributions are, in fact, common within star-forming regions and clusters in the Galaxy (e.g., see Fig. 7 of Amard et al.~2019 and references therein).  Similar to the ONC, the slightly older open cluster, hPer ($\sim13$~Myr), also displays a clear bimodal period distribution, with peaks at $0.9$ and $7$~days in the mass range of 0.3--1.4 $M_\odot$ (Moraux et al.~2013). At young ages the bimodal distribution is firmly established up to $\sim1.5$~\msun\ and there is tentative evidence that this behaviour extends to higher stellar masses (up to at least $4$~\msun - Sabbi et al.~2020).  

In the present work, we suggest that a bimodal spin rate distribution may persist up to mass of $\sim5$~\msun\ in young (few Myr) clusters.  In pre-main sequence (PMS) stars, the radius of a star (at a given age and metallicity) is a  function of its mass.  For a fixed rotation period, the surface rotational velocity, \vrot\, will thus be a  function of mass.  However, at a given mass (within some small tolerance) a bimodal period distribution would be expected to translate to a bimodal \vrot\ for main sequence stars.  In the following sections, we show that a bimodal spin distribution at a few Myr is expected to persist in main sequence stars with ages up to $\sim1.5$~Gyr.

\subsection{Stellar Models with Rotation - Low mass stars ($\lesssim2$~\msun)} 
%\subsubsection{Low mass stars ($\lesssim2$~\msun)}
We have developed this point using PMS models we computed with the same input micro- and macro-physics as in Amard et al. (2019). These models predict the evolution of internal and surface rotation under the action of meridional circulation and shear turbulence (following the Zahn-Maeder formalism and Mathis et al.~2018 for anisotropic turbulence), extraction of angular momentum by magnetized stellar winds (following Matt et al.~2015), disc coupling (i.e. disc locking) in the earliest phases,  and secular evolution. \footnote{Note that A-type stars and earlier lose almost no angular momentum (Matt et al.~2015) as they are not expected to be able to generate a large scale magnetic field through a convective dynamo within their external convective envelope. Although this seems to be mostly the case (Royer \& Zorec~2011, for rapidly rotating A-type stars - although see Wolff et al.~2004 for a different conclusion in PMS stars), it might not be an accurate model for A stars hosting a fossil magnetic field (e.g. Auri{\`e}re et al.~2007; Cantiello \& Braithwaite~2019; Villebrun et al.~2019)}

\begin{figure}
\includegraphics[width=8.5cm]{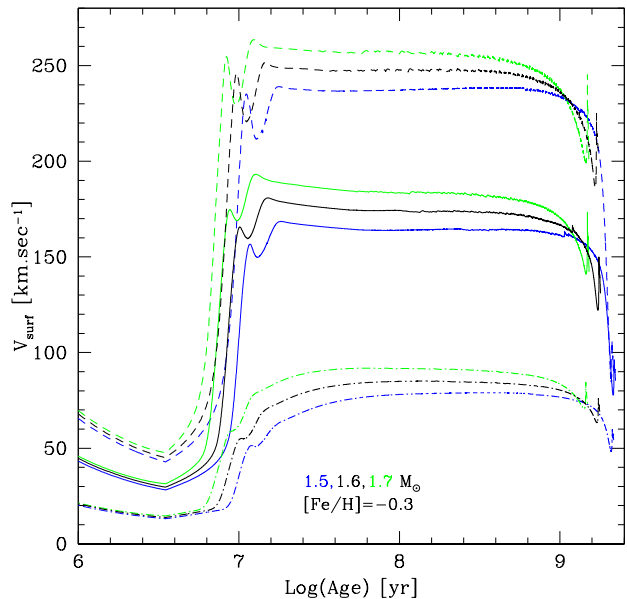}
\caption{The evolution of the surface velocity, V$_{\rm surf}$ from the PMS to the MS turnoff for different masses ($1.5$, $1.6$ and $1.7$~\msun\ shown as blue, black and green, respectively) and different initial rotation rates (periods of $2.3$, $3.5$ and $7.5$~days shown as dashed, solid and dash-dotted, respectively).  All models have a disc lifetime of $3.6$~Myr.  We see that for stars in this mass range (i.e., those near the main sequence turn-off in a $1.5$~Gyr cluster) adopting the observed period distribution of young clusters results in rotational velocities in agreement with observations. Specifically, we find that a bimodal period distribution, as seen in (massive) open clusters is expected to lead to a bimodal rotational distribution of main sequence stars.}
\label{fig:vsurf} \end{figure}

We adopt [Fe/H]=-0.3, appropriate for young and intermediate age clusters in the LMC (e.g. Mucciarelli et al.~2008). To test our suggested model, we have explored how an initial period of a PMS star, at the time of decoupling with the disc, will translate to \vrot\ on the main sequence as a function of stellar mass for three initial periods, $2.3$, $3.5$ and $7.5$ days. We focus on three different masses with a turnoff age around 1.5~Gyr as estimated for NGC~1846 (1.5, 1.6 and 1.7~\msun).  For all models we assume a disc lifetime of $3.6$~Myr, and the results are shown in Fig.~\ref{fig:vsurf}.
%For the 1.5~\msun\ case we compute two additional models with an initial period of 3.5 days and disc lifetimes of 1.8 and 7.6~Myrs. The results are shown in Fig.~\ref{fig:vsurf} \& \ref{fig:vsurf2}.  

In these models, the rotation rates increase substantially after 3.5~Myr, due to the decreasing moment of inertia during pre-main-sequence contraction. The initial periods chosen thus evolve to \vrot\ values between $\sim60$~km/s and $230$~km/s at the arrival on the main sequence.  For stars with masses greater than $\sim 1.4 M_\odot$, the stellar wind torque prescription in the models predict a small amount of angular momentum loss during the pre-main-sequence, but the angular momentum loss is negligible during the main sequence.  Therefore, the initial distribution of rotation rates first translates toward more rapid rates (due to PMS contraction) and then persists essentially unchanged throughout the main sequence lifetime of $\sim 1.5$~Gyr (i.e., to the age of NGC~1846).  %These results show that the rotation rate distributions in young clusters in the Galaxy are expected to span the range of \vrot\ values that are observed in massive LMC/SMC clusters at an age of $1.5$~Gyr (i.e., $60$ and $150$~km/s in Vsini - Kamann et al.~2020). 
Furthermore, whatever is the distribution of rotation rates at an age of $\sim$3~Myr, the shape of the distribution will persist (simply translated to higher rotation velocities) throughout the main-sequence lifetime.  Notably, if the initial rotation rate distribution is bimodal during the PMS, we would expect to then observe bimodal \vrot\ distributions in stars (with masses above $1.5$~\msun) on the MSTO and main sequence for clusters with ages between $\sim10$~Myr and $\sim1.5-2$~Gyr.

%{\bf These models show that an initial bimodal period distribution for PMS stars within a cluster is expected to evolve into a bimodal rotational distribution once the stars reach the main sequence.  Furthermore, after the stars reach the main sequence, at least for stars above $1.5$~\msun\, they would be expected to keep this bimodal rotational distribution throughout their main sequence lifetime.  In terms of applicability to young and intermediate age clusters, we would expect to then observed bimodal \vrot\ distributions for stars (with masses above $1.5$~\msun) on the MSTO and main sequence for clusters with ages between $\sim10$~Myr and $\sim1.5-2$~Gyr.}

% The exact value of the \vrot\ at a given time is dependent on the disc lifetime (as discussed in \S~\ref{sec:origin}).

We note that in the covered mass range, stars with the lower initial period approach or exceed the critical velocity already on the PMS (the fastest models shown in Fig.~\ref{fig:vsurf} evolve between $\sim70$ and $85$\% of critical velocity). How such stars evolve is not clear at present, as the models cannot follow their evolution as they approach the critical velocity.  They are likely to lose mass and angular momentum but remain as rapid rotators.  The observed period distribution in the ONC implies that many of the rapidly rotating PMS stars will get close to or achieve critical rotation (e.g., Amard et al.~2019).  Hence, the observed period distribution of stars in open clusters already suggests that older clusters (assuming they had similar initial distributions) should contain a significant fraction of stars near the critical rotation limit, consistent with observations of massive clusters in the LMC (c.f., Bastian et al. 2017; Milone et al.~2018).

%So far we have just translated the observed period distribution of PMS stars in young open clusters.  To first order we see that the observed period distribution in a cluster like hPer (with peaks in the period distribution of $\sim0.9$ and $\sim7$ days in the mass range of 0.4 to 1.4~\msun - Moraux et al. 2013), is expected to evolve into a population of stars with \vrot\ peaks that are similar to those observed in massive intermediate age clusters.  

Rotation near the critical rate modifies the structural evolution of stars, shifting the location of the main sequence on the HR diagram.
Thus, a young cluster with a bimodal period distribution in its PMS stars would be expected to show a split/dual main sequence once these stars reached the main sequence.  At a given mass, i.e. luminosity in a colour-magnitude diagram, the bimodal \vrot\ distribution would manifest as a splitting, with the red main sequence corresponding to the slow rotators and the blue main sequence corresponding to the rapid rotators (e.g., D'Antona et al. 2015).  The fraction of slow to fast rotators will then be reflected in the fraction of red to blue main sequence stars (at a given magnitude) and be directly related to the fraction of long and short period PMS stars when the cluster was young.

\begin{figure}
\includegraphics[width=8.5cm]{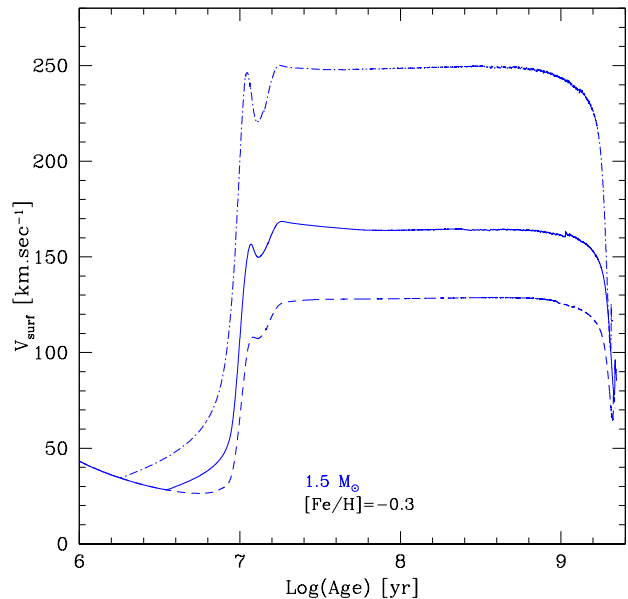}
\caption{The same as Fig.~\ref{fig:vsurf} but only showing the $1.5$~\msun\ model with three different disc lifetimes, 1.8, 3.6, and 7.2~Myr, shown as dash-dotted, solid and dashed lines, respectively.  Longer disc lifetimes result in slower rotational velocities on the main sequence, and vice-versa.}
\label{fig:vsurf2} \end{figure}

\subsection{The Origin of the Bimodal Period Distribution}
\label{sec:origin}

Stars on the PMS are still contracting, so unless a breaking mechanism is able to slow the star down, it will rotate faster as it approaches the main sequence.  
For low-mass stars, it is thought that a magnetic interaction between the star and its accretion disk is able to remove enough angular momentum and prevent significant spin-up, while the disk is present (Camenzind 1990; K\"onigl 1991; Matt \& Pudritz 2005; Bouvier et al.~2014).  
Indeed, there is evidence that within a given young cluster, stars without discs rotate more rapidly than stars with discs (e.g., Edwards et al.~1993; Rebull et al.~2006; Cieza \& Baliber~2007).  
In models of rotational evolution (such as those in Fig.~\ref{fig:vsurf}), the concept of a magnetic star-disk interaction is normally simplified into an assumption called ``disk locking'' (e.g., Bouvier et al.~1993; Bouvier, Forestini, Allain 1997; Rebull et al.~2004; Bouvier et al.~2014).  
Disk locking assumes the rotation period of a forming star is held fixed at some initial spin period, for a paramterized amount of time, which is assumed to correspond to the lifetime of the accretion disk.  
In Fig.~\ref{fig:vsurf}, the disk-locking timescale is fixed at 3.6~Myr, for stars with three different initial spin periods.  Fig.~\ref{fig:vsurf2} shows the evolution of a 1.5~$M_\odot$ star with a single intial spin period of 3.5 days, but three different disk locking times, 1.8, 3.6, and 7.2~Myr.  
This figure demonstrates that a range of spin rates at an age of a few Myr could arise from stars having a range of disk lifetimes, where faster rotators are those whose disks were cleared at earlier ages.
Vasconcelos \& Bouvier~(2015) have run a series of Monte Carlo simulations of low mass stars that include disc-locking for stars with significant accretion rates from their circumstellar discs. 
The authors find that an initially bimodal disc distribution (with and without discs) naturally results in a bimodal period distribution.
Disk locking, or more generally magnetic star-disk interaction, is often assumed to apply to low-mass stars ($<2$~\msun), but we discuss their application to higher masses in \S~\ref{sec:intermediate}.

%Figure~\ref{fig:vsurf2} demonstrates that if the discs cleared earlier it would result in faster \vrot\ values and later disc clearing timescales lead to slower \vrot\ values.   

Hence, it is conceivable that the rotational distribution of stars within a cluster is set in the first few Myr, regulated by whether stars are able to retain their accretion discs, and if so, for how long.  The stellar density and high photoionisation rate within a young massive cluster can be a harsh environment for discs around stars to survive (e.g., Clarke~2007; de Juan Ovelar et al.~2012;  Vincke et al.~2015).  If stars in massive clusters have a higher rate of disc destruction than in lower-mass cluster or associations, then we might expect a larger fraction of rapidly rotating stars within them, which appears to be borne out in observations (e.g., Bastian et al.~2017).  Conversely, stars that are born in a looser association, or far from the nearest ionising high mass stars, would be expected to retain a higher fraction of their discs, leading to more slowly rotating stars (see also Roquette et al.~2017).  Since, high mass dense clusters make up a minority of star-formation in a galaxy (e.g., Bressert et al.~2010; Johnson et al. 2016), this would explain the lower rate of rapidly rotating stars in the field, relative to high mass, dense, clusters (e.g., Strom, Wolff, \& Dror~2005; Huang \& Gies~2008; Bastian et al.~2017).

Additionally, binaries may be expected to destroy discs around the individual stars (e.g., Cesaroni et al.~2007).  Kraus et al.~(2012) observed that stars with binary companions within 40~AU showed a much smaller disk frequency (implying a disk dispersal time $\la 1$~Myr), whereas wider binaries or single stars more likely retained their disks for $3-5$~Myr.  Hence, if the binary fraction has a higher primordial value in dense clusters than in looser associations, this could also play a significant role in destroying the discs, leading to an altered rotational distribution.

However, we emphasise that the main suggestion of this paper is that young or embedded open clusters have a bimodal period distribution amongst their intermediate-mass PMS stars, similar to what is sometimes observed for low mass stars.  This distribution is expected to develop into a bimodal rotational distribution once the stars reach the main sequence (and main sequence turnoff).  Whether disc destruction is responsible for the observed period bimodality at young ages or not is still a matter of debate and a rich avenue for future work.

\subsection{Applicability to Intermediate Mass Stars ($2-5$\msun)}
\label{sec:intermediate}

We do not have much information about the PMS rotation rate distributions for intermediate mass stars, so we do not know whether they are often bimodal.  Furthermore, it is not clear if there is a magnetic star-disk interaction operating in the same way as for low-mass stars (Rosen et al.~2012).

The main problem in extrapolating the models above, which invoke disc-locking as a regulator of the angular momentum of stars in the PMS phase, to intermediate-mass and massive stars is that magnetic fields are detected in less than 10\% of such stars (Grunhut et al.~2017). The mass above which magnetic fields become elusive ($\sim1.5$\msun) coincides with the mass above which main-sequence (MS) stars do not have a convective envelope anymore, which is consistent with the idea that magnetic fields of low-mass stars originate from a convective dynamo (Brun \& Browning 2017).
However, intermediate-mass and massive stars are thought to go through significant convection during their pre-MS phase, due to the low surface temperatures on the Hayashi limit (e.g.~Bernasconi, \& Maeder~1996; Haemmerl{\'e} et al.~2019). A convective dynamo could therefore drive a magnetic field in pre-MS intermediate stars before vanishing once the star has contracted enough to become radiative.

The pre-MS evolution of intermediate-mass and massive stars differs qualitatively from that of low-mass stars, mostly due to the role of accretion. While low-mass stars terminate accretion early in the pre-MS contraction, with a fully convective structure at the top of the Hayashi line, stars with masses $\gtrsim2$~\msun\ do not go through a proper Hayashi phase, because a radiative core forms or is already formed at the end of accretion (Stahler~1983; Palla \& Stahler~1990).  As the star further contracts, this core grows in mass until the convective envelope disappears.

When extrapolating disc-locking to stars with masses $\gtrsim2$~\msun\ we must consider the coupling between the star and a residual disc, after the main accretion phase.  During the main accretion phase, the accretion rates are so high that the stellar magnetosphere (assuming the stars are magnetised) will be crushed onto or close to the stellar surface, and the magnetic star-disk interaction is not able to extract significant angular momentum  (e.g., Rosen et al.~2012).  After the main accretion phase, when the accretion rates decrease, it might be possible for the star-disk interaction to play a role for intermediate mass stars, if the phase lasts sufficiently long or the stars are sufficiently magnetised (Rosen et al.~2012).

% Stellar magnetic fields can transport angular momentum only towards the Alfv\'en radius, and additional mechanisms are required to extract this angular momentum to larger scales, for accretion to proceed.  Moreover, stellar magnetic fields are not expected to be strong enough to funnel efficiently the gas of massive accretion discs (Rosen et al.~2012), so that efficient coupling can occur only after the main accretion phase.

Thus, disc-locking requires the convective dynamo to be still efficient at end the accretion phase, i.e. the convective envelope must still be present when the star reaches its final mass.
For relevant accretion histories, this condition is satisfied for stars with masses up to $\sim5$~\msun\ (Haemmerl{\'e} et al.~2019), at least at solar metallicity, which is approximately the mass up to which the split main sequence has been observed to in young stellar clusters (e.g., Milone et al.~2018).
Note that if fossil fields should survive until the pre-MS phase, magnetic coupling could be ensured for stars of any masses, although this  would concern however a very small fraction of OB stars (only $\sim 7 \%$ have surface magnetic fields; e.g. Keszthelyi et al. 2020 and references therein).

\subsection{Further Evolution After the Main Sequence is Reached}

After the PMS stage stars will not necessarily keep their rotation rate throughout the time on the main sequence, depending on magnetic braking. In the mass range considered, the braking is not expected to be efficient with the current prescriptions as those adopted in our models, and as inferred by observations (e.g. Zorec \& Royer~2012).  Hence, we would expect stars with masses from $\sim1.5-1.7$~\msun\ to retain their initial rotational velocities throughout their main sequence lifetimes (e.g., Georgy et al. 2019).  However, stars with masses below $\sim1.5$~\msun\ generate their own magnetic fields which brake the stars, causing them to eventually become slow rotators (Kraft~1967).  This is evident in the period distribution of stars in open clusters with ages from a few Myr to a few Gyr, where the fraction of fast rotators decreases (at higher masses faster than at lower masses, see Fig. 7 in Amard et al. 2019).  

In fact, this braking from rapidly rotating to slowly rotating near the $\sim1.5$~\msun\ transition, where higher mass stars are still rapidly rotating and lower mass stars have been significantly braked has recently been directly observed in the $1.5$~Gyr clusters, NGC~1846 (Kamann et al. 2020).

\section{Discussion}
\label{sec:discussion}

\subsection{Summary}
We have investigated the origin of the observed dual/split main sequences and observed bimodal rotational distributions in massive stellar clusters in the LMC/SMC, with ages between $\sim50$~Myr (e.g., Milone et al. 2018) and $\sim1.5$~Gyr (Kamann et al.~2020).  By looking at the period distribution of pre-main sequence stars in nearby star-forming regions, we note that it tends to be bimodal largely independent of stellar mass (from $0.4-1.5~\msun$ with evidence that it continues to $>4$~\msun).  This implies a wide range of rotational velocities (which are a convolution of the period and the stellar radius which is mass dependent) to be present within a cluster.  However, when stars are on the main sequence, stars with comparable masses will have similar luminosities, and so at a given luminosity, we may expect to observe a splitting due to the bimodal rotational distribution at/near that mass.

% {\bf We have computed models accounting for state-of-the-art prescriptions for the internal transport and the extraction of angular momentum.  We showed that if we start with an initial bimodal period distribution (i.e., at ages $<2-3$~Myr) we end up with a bimodal period (and \vrot) distribution already at the age of hPer ($13$~Myr) which remains for the duration of the PMS and MS lifetimes of the stars within the cluster (for a fixed stellar mass and disc lifetime).  Hence, we conclude that the bimodal Vsini distribution (and split/dual main sequences) observed in clusters with ages between $\sim10$~Myr and $\sim1.5$~Gyr is already set in the early stages of a cluster's lifecycle, when the stars are still on the PMS. } 

We have computed models accounting for the evolution of stellar structure and angular momentum.  We demonstrated that if cluster stars with 1.5-5 $M_\odot$ have an initially bimodal period (and \vrot) distribution at ages $<2-3$~Myr, this bimodality will persist for the duration of the PMS and main-sequence lifetime.
A bimodal stellar rotational distribution could explain an eMSTO as well as a dual/split MS (e.g., Bastian \& de Mink~2009; D'Antona et al.~2015).  
Hence, we suggest that the origin of the split main sequence in young ($\sim10$~Myr) massive clusters, as well as the observed bimodal Vsini distributions in intermediate aged ($\sim1.5$~Gyr) clusters, is (at least partially) due to a bimodal spin distribtuion that is already present at ages of a few Myr (or earlier).

It is currently unclear whether such a distribution alone can quantitatively match all aspects of the observed CMDs, with some studies finding that in addition to a rotational distribution an age spread may also be required (e.g., Goudfrooij et al.~2017).  However, it should be noted that the treatment of rotation into stellar evolutionary tracks still suffers some uncertainties (even on the MS), related in particular to the approximations required to treat sophisticated processes like turbulence (e.g. Ekstr{\"o}m et al.~2018, Mathis et al.~2018).  Hence, it may not be surprising that the models do not provide perfect fits to observations.  Further benchmarking of key parameters against detailed observations (i.e., colours, magnitudes and measured rotation rates) is needed (i.e., Gossage et al.~2019), especially given the strong constraints against significant age spreads within clusters that are model independent (see Bastian \& Lardo~2018 for a recent review).   

In constructing our model we have extrapolated the period distribution of relatively low mass stars (mainly with $m < 1.5$~\msun) in young open clusters into the mass range $1.5-1.7$~\msun.  
Observations of resolved high mass clusters, like those of Sabbi et al.~(2020) for Westerlund~2, should be able to show whether more massive stars (up to $\sim4$~\msun) also display a bimodal period distribution.
The origin of a bimodal period distribution amongst PMS stars in young clusters is still not entirely clear, but it could be due to a star-disc interaction during the PMS lifetime of the star.  
Note that our adopted model of disc regulated angular momentum in stars was primarily developed for relatively low mass stars ($\lesssim1.5$~\msun).  
It is not yet clear whether the protoplanetary disc can play the same role for higher mass stars.
We note, however, that observations of the eMSTO and split main sequence phenomena do not appear to show any discontinuities above $\sim1.5$~\msun, potentially suggesting a common origin across all stellar masses.
Observations of young star forming regions that span this boundary can shed more light on whether there is a mass dependence in the period distribution within these regions and potentially the role of different mechanisms in controlling the angular momentum of young stars.

One possibility, discussed in the present work, is that a PMS period bimodality is due to whether a star is able to retain its protoplanetary disc (which would allow it to be disc-braked, resulting in a slowly rotating star) or loses its disc due to interactions and/or photoionisation from nearby high-mass stars (resulting in rapidly rotating stars).    We would then expect a dependence on the environment, which would control when and what fraction of stars lose their discs.  Hence, the actual velocity peaks in the distributions are likely to vary from cluster to cluster.  
The final rotation rate of the stars is sensitive to the time at which the disc is lost, so we may not expect a clear bimodality in the resulting \vrot\ distribution unless the discs are lost in somewhat discrete epochs.  On the other hand, if there is convergent evolution in the rotation rate, i.e., all disc cleared stars evolve close to critical rotation resulting in a similar final rotation rate, this may explain the bimodality.  Perhaps studies of young star forming regions and the period distributions of their stars (in the mass range 1.5-5 $M_\odot$) as a function of age and disc fraction could disentangle the various mechanisms.
Measuring the Vsini distribution in a sample of clusters that span a wide range of ages and densities will also allow many of the assumptions and predictions of the presented model to be tested and shed light on the possible origins of bimodal distributions.

% \subsection{Caveats}

% \begin{itemize}
% \item As mentioned above, models for the evolution of angular momentum in young stars regulated by the protoplanetary disc have mainly been focussed on lower mass stars ($\lesssim1.5$~\msun).  It is not clear if disc-locking is relevant for stars more massive than this limit and what controls the final angular moment within these stars.  We note, however, that observations of the eMSTO and split main sequence phenomena do not appear to show any discontinuities above $\sim1.5$~\msun, potentially suggesting a common origin across all stellar masses.

% \item For the stellar mass range focussed on in this paper, the final rotation rate of the stars is sensitive to the time at which the disc is lost, hence we may not expect a clear bimodality in the resulting \vrot\ distribution unless the discs are lost in somewhat discrete epochs.  On the other hand, if there is convergent evolution in the rotation rate, i.e., all disc cleared stars evolve close to critical rotation resulting in a similar final rotation rate, this may explain the bimodality.  Perhaps studies of young star forming regions and the period distributions of their stars as a function of age and disc fraction could disentangle the various mechanisms.

% \end{itemize}

\subsection{Predictions}

While the model presented here has a number of free parameters and hopefully will open the door for future theoretical/numerical investigations, it already makes a number of predictions that can be tested observationally.  Below we outline a few of them:

\begin{itemize}

\item If the bimodality is set up due to the destruction of discs from dynamical interactions and photoinisation, both of which increase for more massive/denser clusters as well as towards the centre of the cluster, the models predicts that the fraction of rapid rotators should increase in both cases.  If true, we would expect the radial profile of the rapid rotators within a cluster to be more centrally concentrated than the slower rotators.  This is indeed the case for the $\sim200$~Myr, LMC cluster, NGC~1866 (Milone et al.~2018) although in the $\sim300$~Myr, LMC cluster NGC~1856 the rapid rotators only show a slight preference towards the cluster centre (Li et al. 2017).  We note, however, that relaxation will wash out any primordial spatial differences between the populations over time.  Additionally, the fraction of Be stars (thought to be rapid rotators near critical rotational velocity) appears to be higher in more massive / denser clusters (e.g., Bastian et al.~2017; Milone et al. 2018).

Similarly, clusters whose stars are able to retain their discs for longer (lower density clusters, and clusters without strong photoionisation sources - i.e., without O-stars) should have fewer rapidly rotating stars.

\item As opposed to the 'binary interaction' model which is expected to take 10s of Myr before binary interactions have a strong influence on the fraction of rapid/slow rotators, the present model predicts that the bimodal period distribution (and Vsini distribution if restricted to a small mass range) should be in place within a few Myr of the cluster's birth.  An excellent location to test this is in R136, a $\sim3$~Myr massive cluster in the LMC.  We note that in the larger 30~Doradus region (although focussed on R136), Dufton et al.~(2013) have found a bimodal rotational distribution in B-stars with (de-projected) rotational velocity peaks at $\sim60$ and $290$~km/s.  Whether the lower mass PMS stars within this cluster have a bimodal distribution is an excellent area for future study.

\item The period distribution appears to be largely independent of stellar mass in clusters with ages of a few Myr, and the conversion from period to Vsini is mass (radius) dependent.  Hence, we would expect the Vsini value of the rapid rotation peak of the distribution should increase towards higher stellar mass (or younger ages if looking consistently at the main sequence turnoff part of the CMD).  We note that the observations of B-stars in 30~Doradus by Dufton et al.~(2013 - discussed above) appear to fit this trend.

{\bf}
\end{itemize}

%\vspace{-0.7cm}

\section*{Acknowledgments}

The authors thank Franca D'Antona, Norbert Langer and Nathan Mayne for insightful discussions. NB gratefully acknowledges financial support from the Royal Society (University Research Fellowship).  NB and SK gratefully acknowledges financial support from the European Research Council (ERC-CoG-646928, Multi-Pop).  LA and SPM gratefully acknowledges funding from the European Research Council (ERC 682393, AWESoMeStars). CC and LH acknowledge support from the Swiss National Science Foundation (SNF) for the project 200020-169125 "Globular cluster archeology".

\bsp
\label{lastpage}

\vspace{-0.5cm}
\begin{thebibliography}{99}

\bibitem[Amard et al.(2019)]{2019A&A...631A..77A} Amard, L., Palacios, A., Charbonnel, C., et al.\ 2019, A\&A, 631, A77

\bibitem[Auri{\`e}re et al.(2007)]{2007A&A...475.1053A} Auri{\`e}re, M., Wade, G.~A., Silvester, J., et al.\ 2007, A\&A, 475, 1053

\bibitem[Bastian \& de Mink(2009)]{BastianDeMink09} Bastian, N., \& de Mink, S. E. 2009, MNRAS, 398, L11

\bibitem[Bastian et al.(2017)]{2017MNRAS.465.4795B} Bastian, N., Cabrera-Ziri, I., Niederhofer, F., et al.\ 2017, MNRAS, 465, 4795

\bibitem[Bastian et al.(2018)]{2018MNRAS.480.3739B} Bastian, N., Kamann, S., Cabrera-Ziri, I., et al.\ 2018, MNRAS, 480, 3739

\bibitem[Bastian \& Lardo(2018)]{BL18} Bastian, N. \& Lardo, C.~2018, ARA\&A, 56, 83

\bibitem[Beasor et al.(2019)]{2019MNRAS.486..266B} Beasor, E.~R., Davies, B., Smith, N., et al.\ 2019, MNRAS, 486, 266

%\bibitem[Beccari et al.(2017)]{2017A&A...604A..22B} Beccari, G., Petr-Gotzens, M.~G., Boffin, H.~M.~J., et al.\ 2017, A\&A, 604, A22

\bibitem[Bernasconi, \& Maeder(1996)]{1996A&A...307..829B} Bernasconi, P.~A., \& Maeder, A.\ 1996, A\&A, 307, 829

\bibitem[Bressert et al.(2010)]{2010MNRAS.409L..54B} Bressert, E., Bastian, N., Gutermuth, R., et al.\ 2010, MNRAS, 409, L54

\bibitem[Bouvier et al.(1993)]{1993A&A...272..176B} Bouvier, J., Cabrit, S., Fernandez, M., et al.\ 1993, A\&A, 272, 176

\bibitem[Bouvier et al.(2014)]{2014prpl.conf..433B} Bouvier, J., Matt, S.~P., Mohanty, S., et al.\ 2014, Protostars and Planets VI, 433

\bibitem[Bouvier et al.(1997)]{bouvier97} Bouvier, J., Forestini, M., Allain, S. 1997, A\&A, 326, 1023

\bibitem[Brun \& Browning(2017)]{brun17} Brun, A.~S., Browning, M.~K. 2017, Living Rev. Sol. Phys., 14, 4

\bibitem[Camenzind(1990)]{camenzind90} Camenzind, M. 1990, RvMA, 3, 234

\bibitem[Cesaroni et al.(2007)]{2007prpl.conf..197C} Cesaroni, R., Galli, D., Lodato, G., et al.\ 2007, Protostars and Planets V, 197

\bibitem[Cieza, \& Baliber(2006)]{cieza06} Cieza, L., \& Baliber, N.\ 2006, ApJ, 649, 862

\bibitem[Cieza, \& Baliber(2007)]{2007ApJ...671..605C} Cieza, L., \& Baliber, N.\ 2007, ApJ, 671, 605

\bibitem[Clarke(2007)]{2007MNRAS.376.1350C} Clarke, C.~J.\ 2007, MNRAS, 376, 1350

\bibitem[Cantiello \& Braithwaite(2019)]{2019ApJ...883..106C} Cantiello, M., \& Braithwaite, J.\ 2019, ApJ, 883, 106

\bibitem[D'Antona et al.(2015)]{2015MNRAS.453.2637D} D'Antona, F., Di Criscienzo, M., Decressin, T., et al.\ 2015, MNRAS, 453, 2637

\bibitem[D'Antona et al.(2017)]{2017NatAs...1E.186D} D'Antona, F., Milone, A.~P., Tailo, M., et al.\ 2017, Nature Astronomy, 1, 0186

\bibitem[de Juan Ovelar et al.(2012)]{2012A&A...546L...1D} de Juan Ovelar, M., Kruijssen, J.~M.~D., Bressert, E., et al.\ 2012, A\&A, 546, L1

\bibitem[Dufton et al.(2013)]{2013A&A...550A.109D} Dufton, P.~L., Langer, N., Dunstall, P.~R., et al.\ 2013, A\&A, 550, A109

\bibitem[Dupree et al.(2017)]{dupree17} Dupree, A.~K., Dotter, A., Johnson, C.~I., et al.\ 2017, ApJL, 846, L1 

\bibitem[Edwards et al.(1993)]{edwards93} Edwards, S. et al.~1993, AJ, 106, 372

\bibitem[Ekstr{\"o}m et al.(2018)]{2018MmSAI..89...50E} Ekstr{\"o}m, S., Meynet, G., Georgy, C., et al.\ 2018, Memorie della Soci-
eta Astronomica Italiana, 89, 50

%\bibitem[Jeffries et al.(2011)]{2011MNRAS.418.1948J} Jeffries, R.~D., Littlefair, S.~P., Naylor, T., et al.\ 2011, MNRAS, 418, 1948

\bibitem[Giesers et al.(2019)]{2019A&A...632A...3G} Giesers, B., Kamann, S., Dreizler, S., et al.\ 2019, A\&A, 632, A3

\bibitem[Georgy et al.(2019)]{2019A&A...622A..66G} Georgy, C., Charbonnel, C., Amard, L., et al.\ 2019, A\&A, 622, A66

\bibitem[Gossage et al.(2019)]{2019ApJ...887..199G} Gossage, S., Conroy, C., Dotter, A., et al.\ 2019, ApJ, 887, 199

\bibitem[Grunhut et al.(2017)]{2017MNRAS.465.2432G} Grunhut, J.~H., Wade, G.~A., Neiner, C., et al.\ 2017, MNRAS, 465, 2432

\bibitem[Haemmerl{\'e} et al.(2019)]{2019A&A...624A.137H} Haemmerl{\'e}, L., Eggenberger, P., Ekstr{\"o}m, S., et al.\ 2019, A\&A, 624, A137

\bibitem[Herbst et al.(2001)]{herbst01} Herbst, W., Bailer-Jones, C.~A.~L., Mundt, R. 2001, ApJ, 554, L197

\bibitem[Herbst et al.(2002)]{2002A&A...396..513H} Herbst, W., Bailer-Jones, C.~A.~L., Mundt, R., et al.\ 2002, A\&A, 396, 513

\bibitem[Huang \& Gies(2008)]{2008ApJ...683.1045H} Huang, W., \& Gies, D.~R.\ 2008, ApJ, 683, 1045

\bibitem[Hurley et al.(2007)]{2007ApJ...665..707H} Hurley, J.~R., Aarseth, S.~J., \& Shara, M.~M.\ 2007, ApJ, 665, 707

\bibitem[Hut et al.(1992)]{1992PASP..104..981H} Hut, P., McMillan, S., Goodman, J., et al.\ 1992, PASP, 104, 981

\bibitem[Irwin \& Bouvier(2009)]{irwin09} Irwin, J., Bouvier, J 2009, Proc. IAU Symp. 258, The Ages of Stars, Eds. Mamajek, E.~E., Soderblom, D.~R., Wyse, R.~F.~G., p. 363

\bibitem[Johnson et al.(2016)]{2016ApJ...827...33J} Johnson, L.~C., Seth, A.~C., Dalcanton, J.~J., et al.\ 2016, ApJ, 827, 33

\bibitem[Kamann et al.(2018)]{2018MNRAS.480.1689K} Kamann, S., Bastian, N., Husser, T.-O., et al.\ 2018, MNRAS, 480, 1689

%\bibitem[Kamann et al.(2020)]{temp} Kamann, S., Bastian, N. et al.\ 2020, MNRAS, in press
\bibitem[Kamann et al.(2020)]{2020MNRAS.492.2177K} Kamann, S., Bastian, N., Gossage, S., et al.\ 2020, MNRAS, 492, 2177


%\bibitem[Littlefair et al.(2011)]{2011MNRAS.413L..56L} Littlefair, S.~P., Naylor, T., Mayne, N.~J., et al.\ 2011, MNRAS, 413, L56

%\bibitem[Keszthelyi et al.(2020)]{temp2} Keszthelyi, Z. et al.~2020, MNRAS, in press (arXiv: 2001.06239)

\bibitem[Keszthelyi et al.(2020)]{2020MNRAS.493..518K} Keszthelyi, Z., Meynet, G., Shultz, M.~E., et al.\ 2020, MNRAS, 493, 518

\bibitem[Kraft(1967)]{1967ApJ...150..551K} Kraft, R.~P.\ 1967, ApJ, 150, 551

\bibitem[K\"onigl(1991)]{konigl91} K\"onigl, A. 1991, ApJ, 370, L39

\bibitem[Kraus et al.(2012)]{kraus12} Kraus, A.~L. Ireland, M.~J., Hillenbrand, L.~A., and Martinache, F. 2012, ApJ, 745, 19

\bibitem[Lamm et al.(2005)]{Lamm05} Lamm, M.~H., Mundt, R., Bailer-Jones, C.~A.~L., Herbst, W. 2005, A\&A, 430, 1005

\bibitem[Li et al.(2017)]{2017ApJ...834..156L} Li, C., de Grijs, R., Deng, L., et al.\ 2017, ApJ, 834, 156

\bibitem[Mackey \& Broby Nielsen(2007)]{MackeyBrobyNielsen07}  Mackey, A.~D., \& Broby Nielsen, P.\ 2007, MNRAS, 379, 151 

\bibitem[Marino et al.(2018)]{2018AJ....156..116M} Marino, A.~F., Przybilla, N., Milone, A.~P., et al.\ 2018, AJ, 156, 116

\bibitem[Mathis et al.(2018)]{2018A&A...620A..22M} Mathis, S., Prat, V., Amard, L., et al.\ 2018, A\&A, 620, A22

\bibitem[Matt \& Pudritz(2005)]{matt15} Matt, S.~P., Pudritz, R.~E. 2005, Apj, 632, L135

\bibitem[Matt et al.(2015)]{2015ApJ...799L..23M} Matt, S.~P., Brun, A.~S., Baraffe, I., et al.\ 2015, ApJL, 799, L23

\bibitem[Milone et al.(2012)]{2012A&A...540A..16M} Milone, A.~P., Piotto, G., Bedin, L.~R., et al.\ 2012, A\&A, 540, A16

\bibitem[Milone et al.(2016)]{2016MNRAS.458.4368M} Milone, A.~P., Marino, A.~F., D'Antona, F., et al.\ 2016, MNRAS, 458, 4368 

\bibitem[Milone et al.(2018)]{2018MNRAS.477.2640M} Milone, A.~P., Marino, A.~F., Di Criscienzo, M., et al.\ 2018, MNRAS, 477, 2640

\bibitem[Moraux et al.(2013)]{2013A&A...560A..13M} Moraux, E., Artemenko, S., Bouvier, J., et al.\ 2013, A\&A, 560, A13

\bibitem[Mucciarelli et al.(2008)]{2008AJ....136..375M} Mucciarelli, A., Carretta, E., Origlia, L., et al.\ 2008, AJ, 136, 375

\bibitem[Palla, \& Stahler(1990)]{1990ApJ...360L..47P} Palla, F., \& Stahler, S.~W.\ 1990, ApJL, 360, L47

\bibitem[Ram{\'\i}rez-Agudelo et al.(2015)]{2015A&A...580A..92R} Ram{\'\i}rez-Agudelo, O.~H., Sana, H., de Mink, S.~E., et al.\ 2015, A\&A, 580, A92

\bibitem[Rebull et al.(2004)]{2004AJ....127.1029R} Rebull, L.~M., Wolff, S.~C., \& Strom, S.~E.\ 2004, AJ, 127, 1029

\bibitem[Rebull et al.(2006)]{rebull06} Rebull, L.~M., Stauffer, J.~R., Megeath, S.~T., Hora, J.~L., and Hartmann, L. 2006, ApJ, 646, 297

\bibitem[Roquette et al.(2017)]{2017A&A...603A.106R} Roquette, J., Bouvier, J., Alencar, S.~H.~P., et al.\ 2017, A\&A, 603, A106

\bibitem[Rosen et al.(2012)]{2012ApJ...748...97R} Rosen, A.~L., Krumholz, M.~R., \& Ramirez-Ruiz, E.\ 2012, ApJ, 748, 97

\bibitem[Sabbi et al.(2020)]{2020ApJ...891..182S} Sabbi, E., Gennaro, M., Anderson, J., et al.\ 2020, ApJ, 891, 182

\bibitem[Stahler(1983)]{1983ApJ...274..822S} Stahler, S.~W.\ 1983, ApJ, 274, 822

\bibitem[Strom et al.(2005)]{2005AJ....129..809S} Strom, S.~E., Wolff, S.~C., \& Dror, D.~H.~A.\ 2005, AJ, 129, 809

\bibitem[Vincke et al.(2015)]{2015A&A...577A.115V} Vincke, K., Breslau, A., \& Pfalzner, S.\ 2015, A\&A, 577, A115

\bibitem[Vasconcelos \& Bouvier(2015)]{2015A&A...578A..89V} Vasconcelos, M.~J., \& Bouvier, J.\ 2015, A\&A, 578, A89

\bibitem[Villebrun et al.(2019)]{2019A&A...622A..72V} Villebrun, F., Alecian, E., Hussain, G., et al.\ 2019, A\&A, 622, A72

\bibitem[Wang et al.(2020)]{2020ApJ...888L..12W} Wang, C., Langer, N., Schootemeijer, A., et al.\ 2020, ApJL, 888, L12

\bibitem[Wolff et al.(2004)]{2004ApJ...601..979W} Wolff, S.~C., Strom, S.~E., \& Hillenbrand, L.~A.\ 2004, ApJ, 601, 979

\bibitem[Zorec, \& Royer(2012)]{2012A&A...537A.120Z} Zorec, J., \& Royer, F.\ 2012, A\&A, 537, A120


%---------------------------------------------------------
%-----------------------------------------------------



\end{thebibliography}
\end{document}